\documentclass[12pt,preprint]{aastex}

\shorttitle{Tr 16 and Gaia} 
\shortauthors{Davidson et al.}

\begin{document}

\title{Gaia, Trumpler 16, and Eta Carinae} 

%% \revised{xxx} 

\author{Kris Davidson\altaffilmark{1} }  
\author{Greta Helmel\altaffilmark{2} }  
\author{Roberta M. Humphreys\altaffilmark{1} }

\altaffiltext{1}{Minnesota Institute for Astrophysics,   
  116 Church St. SE, Minneapolis, MN 55455 U.S.A.} 
\altaffiltext{2}{Department of Physics and Astronomy, Macalester College, 1600 Grand Ave. St. Paul, MN 55105 U.S.A.}

\keywords{open clusters and associations: individual(Tr 16) - stars: individual(eta Car) - stars: distances}   

\vspace{4mm}  

  (This article is nearly identical to an AAS Research Note: 
  RNAAS 2:133, 2018 Aug 1.  Several minor omissions and ambiguities 
  have been corrected here.) 

\vspace{1mm} 

The star cluster Tr 16 contains four O3-type members and $\eta$ Car -- an 
extraordinary concentration of stars  with $M \gtrsim 50 \, M_\odot$.  Here 
we note that Gaia parallaxes  reveal a discrepancy in $\eta$ Car's distance, 
and also a serious overestimate of cluster membership. For information  
about $\eta$ Car, see various authors' reviews in \citet{dh12}.     

\vspace{1mm} 

We employ Gaia Data Release 2 \citep{br18,lu18}, assuming that the error 
estimates $\sigma_i$ have the characteristics described by those authors.   
We assembled a list of fifty stars which are said to be members of Tr 16 
\citep{nrw73,nrw95,mj93}, all having Gaia parallaxes  
$\varpi_i$ with standard errors $\sigma_i < 0.050$ mas (Fig.\ 1).    
Their weighted average is $\varpi_{av} \approx 0.373$ mas 
and their r.m.s.  $\sigma_i$ is 0.034 mas.  (Each $\varpi_i$ has 
relative statistical weight $1/{\sigma_i}^2$.)  

  \begin{figure} 
  \epsscale{0.6}  
  \plotone{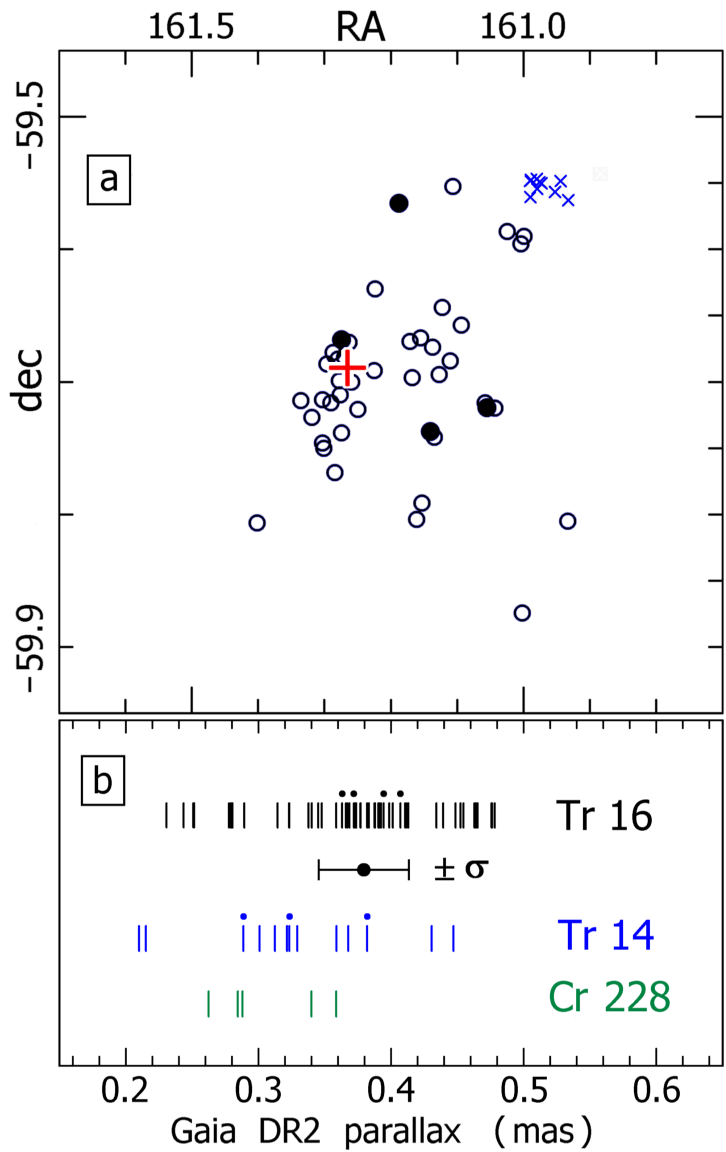} 
  \caption{(a) Apparent positions of stars in the Gaia Tr 16 sample.  
      Filled circles are O3 stars, the large cross is $\eta$ Car, and 
      the small x's belong to Tr 14.  Seven stars with the most 
      discrepant parallaxes have been omitted. 
      (b) Distribution of Gaia parallaxes in Tr 16, Tr 14, and Cr 228. 
      For Tr 16 the entire 50-star sample is included. Dots mark the O3 
      stars, and a  ${\pm}1{\sigma}$ error bar shows the r.m.s. individual 
      $\sigma_i$ in the Tr 16 sample. }    
  \end{figure}

The Gaia data show that many of the 50 stars probably do not belong to 
Tr 16.  This cluster's diameter,  roughly 15 pc, corresponds to a range 
of only 0.002 mas in parallax.  Thus its  distribution of $\varpi_i$ 
values should be indistinguishable 
from a sample with one true parallax and errors consistent with the 
$\sigma_i$'s.  In fact, however, 13 of the 50 stars have 
$|\varpi_i - \varpi_{av}| > 2.5 \, \sigma_i$.   If we model the  
observed $\varpi_i$ distribution as a cluster superimposed on a set of 
stars randomly scattered along the Carina spiral arm, then we find 
$N(\mathrm{nonmembers}) \gtrsim 20$.  O- and B-type stars are very 
numerous along that line of sight, and distance-dependent factors played 
only a small role in the  membership criteria for 
Tr 16.  Hence the sample does not provide 
a robust estimate of the cluster parallax, because there is no good way 
to decide  which stars to include.  The formal error of their average 
is only about $\pm0.005$ mas, but this has little meaning for a 
contaminated sample.

Now consider only the four O3 stars: HD 93205, HD 93250, HD 303308, and 
MJ 257 \citep{nrw95,mj93}.  Here ``O3'' includes proposed subtypes 
O2, O2.5, O3, and O3.5, since the distinctions between them have no 
significance for this discussion.  Only about a dozen O3's are known for 
the entire Galaxy; Tr 16 has four of them within a $0.2^\circ$ circle 
(Fig.\ 1a); and very massive stars tend to occur in groups \citep{hd84}. 
Thus we have clear reasons to postulate that these four belong 
to a physical group, identified long before Gaia.  
If they do, then they give a reliable collective parallax estimate.

The data support the hypothesis.  The four O3 stars' weighted mean 
is $\varpi_{av} = 0.383$ mas, and all of them have 
$|\varpi_i - \varpi_{av}| < 0.8 \, \sigma_i$; \ so their spread in 
$\varpi_i$ is quite small -- indeed, smaller than expected.     Hence 
we propose that the parallax of  Tr 16 is 0.383 $\pm$ 0.017 mas,  
where the r.m.s.\ error estimate is based on the Gaia $\sigma_i$'s 
for these four stars.

Eta Car, which is even more massive than the O3 stars, has long been 
considered a member of Tr 16, see Fig.\ 1a and \citet{nrw95,nrw12}.   
This object is unsuitable for parallax 
measurement,  because it has bright, asymmetric, varying ejecta and 
orbital motion.  But its larger-scale ejecta provide another distance 
method, based on Doppler velocities and proper motions of outward-moving 
condensations \citep{th56,ha92,meab93,kd01}.  For statistical  purposes 
the result is much like a parallax, because the most difficult observables 
are  proper motions, which are proportional to $1/D$.   The 2001 Hubble 
Space Telescope result was $\varpi \approx 0.440 \pm 0.035$ mas,    
and no improvement has been achieved since that time, since the HST 
had the best spatial resolution for this problem.  But our Tr 16 Gaia 
result (see above) disagrees, and is probably better.

{\it Therefore we strongly suspect that $\eta$ Car's distance is close to 
2600 pc, rather than 2300 pc which nearly all recent authors have adopted.} 
Of course this implies a 25\%--30\% increase in luminosity.  In principle 
$\eta$ Car might be unrelated to Tr 16, or the Gaia DR2 results may have 
large systematic errors;  but these possibilities appear 
considerably less likely.

Two other clusters Tr 14 and Cr 228 are often mentioned in connection 
with Tr 16.  Tr 14, a compact group marked in Fig. 1a, contains three more 
O3 stars, but  \citet{nrw95} concluded that it is distinct from Tr 16 because 
its stars are younger.   Gaia data give $\varpi_{av} = 0.327 \pm 0.018$ mas 
for the Tr 14 O3 stars, placing them about $450 \pm 200$ pc beyond Tr 16.  
If this is correct, then Tr 14 has no relation to $\eta$ Car.  Cr 228, 
a large, relatively sparse cluster just below the bottom edge of Fig. 1a, 
does not affect the above discussion.

{\it Acknowledgement:}   
This note employs data from the European Space Agency {\it Gaia} 
mission, https://www.cosmos.esa.int/gaia.

      \vspace{8mm}    %%% ===-=== %%% 

\end{document}